\documentclass{icrc2009}

\usepackage{graphicx}   
\usepackage[font=footnotesize]{subfig} 
\usepackage{fixltx2e}
\usepackage{url}

\newcommand{\shorttitle}[1]%
{\markboth{Proceedings of the 31\MakeLowercase{$^{st}$} ICRC, {\L}\'{o}d\'{z} 2009}{#1} }
\newcommand{\etal}{\MakeLowercase{\textit{et al. }}} 

\begin{document}
\title{Acoustic sensor development for ultra high energy neutrino detection}

\author{\IEEEauthorblockN{Matt Podgorski\IEEEauthorrefmark{2} and Mathieu Ribordy\IEEEauthorrefmark{1}}
                            \\
\IEEEauthorblockA{\IEEEauthorrefmark{1} High Energy Physics Laboratory, EPFL, CH - 1015 Switzerland}
\IEEEauthorblockA{\IEEEauthorrefmark{2}RWTH Aachen, visiting EPFL}}

\shorttitle{Mathieu Ribordy \etal acoustic sensor R\&D}
\maketitle

\begin{abstract}
The GZK neutrino flux characterization would give insights into cosmological source evolution, source spectra and composition at injection from the partial recovery of the degraded information carried by the ultra high energy cosmic rays. 
The flux is expected to be at levels necessitating a much larger instrumented volume ($>$100 km$^3$) than those currently operating. 
First suggested by Askaryan, both radio and acoustic detection techniques could render this quest possible thanks to longer wave attenuation lengths (predicted to exceed a kilometer) allowing for a much sparser instrumentation compared to optical detection technique. \\
We present the current acoustic R\&D activities at our lab developing adapted devices, report on the obtained sensitivies and triangulation capabilities we obtained, and define some of the requirements for the construction of a full scale detector.
  \end{abstract}

 
\section{Introduction}
The IceCube detector \cite{icecube} may well soon identify the first ultra high energy neutrino of cosmogenic origin, following interactions of ultra high energy cosmic rays with the cosmic microwave background \cite{gzk}. Predictions for the cosmogenic neutrino flux, i.e. neutrinos from photo-disintegration, is at levels  of the order of $E dN/dE \sim 10^{-17} \,\mathrm{s}^{-1}\,\mathrm{cm}^{-3}\mathrm{sr}^{-1}$ at $E=10^{18}$ eV, resulting in 0.01 - 1 event / year / km$^3$ in ice \cite{Engel:2001hd}. These predictions strongly depend on the primary cosmic ray composition \cite{Anchordoqui:2007fi}. 
Currently, the situation is uncertain: While the observed correlation of UHE CR sources with the AGN distribution by AUGER \cite{Abraham:2007si} hints toward a light composition (and in this case we lie close to the upper flux predictions), dedicated AUGER composition studies favor a composition turning heavier at UHE \cite{Kampert:2009zz}.
GZK neutrinos are astronomical messengers keeping track of the original CR direction, GZK interactions mostly occur close to the source. In case of the existence of a few UHE cosmic accelerators located close-by (Gpc), the detection of a substantial flux of GZK neutrinos from these directions using a multi-messenger approach would allow the possibility of pinpointing the nature of these CR accelerators.

The characterization of the GZK neutrino flux spectrum, and thus the recovery of the degraded information information carried by UHE CR, would allow the delineation of cosmological source evolution scenarios from source spectrum characteristics.
To fulfill this goal, the event detection rate should be vastly increased. Therefore a much larger volume should be instrumented with an adequate technology for the detection of ultra high energy neutrino interactions.
Two novel detection methods have been proposed, following signatures first discussed by Askaryan \cite{aska:a, aska:c}. An interacting neutrino emits a coherent Cherenkov pulse in the range of 0.1-1 GHz \cite{Zas:1991jv} close to the shower axis and a thin thermoacoustic pancake normal to the shower axis. Both detection techniques are currently exploited by several detectors. In ice, both radio and acoustic emissions have rather large theoretical attenuation lengths \cite{price}. While this has been convincingly demonstrated for the radio emission, it is still a work in progress for the acoustic emission and is one of the main goals for the South Pole Acoustic Test Setup (SPATS) \cite{Boeser:2008bj}. With the data collected by the SPATS array, the sound speeds w.r.t. the depth have been determined with great accuracy, meeting theoretical expectations \cite{Vandenbroucke:2008bs} and S-waves have been found as well. Unknown, however, remains the exact nature of the local source of noise and the exact value of the attenuation length. Newest experimental results hint toward a reduced pressure wave attenuation length on one hand and demonstrate favorable noise level below 10 mPa  on the other hand \cite{Descamps}.

Contrary to salt and water, ice is unique. It allows the detection of three distinct signatures accompanying a neutrino event: Optical Cherenkov light, coherent radio Cherenkov and thermoacoustic emissions, thus firmly establishing the event origin by a strong background reduction.
A possible layout for the hybrid instrumentation of a large volume of order of 100 km$^3$ at the South Pole would consist of strings deployed one kilometer apart down to a depth of 2 km (radio and acoustic attenuation lengths strongly vary with temperature and are decreasing with depth). Given the topologies for the radio \& acoustic emissions, a string should be densely equipped with radio and acoustic devices with an option of supplementing it with PMT devices for optical detection.
$10 - 10^3$ interacting GZK neutrinos in 100 km$^3$ instrumented volume can be expected after 10 years. The cosmogenic spectrum could be characterized (and consequently insights into the underlying physics), provided a high detection efficiency, a deep knowledge of the local source of noise and good energy resolution. Thermoacoustic models and Monte Carlo simulations predict that a signal from a neutrino with an energy $E=10^{18}$ eV will typically have an amplitude of 10 mPa at a distance of one kilometer \cite{Besson:2005re} (to be rescaled for finite attenuation length). To keep a good S/N ratio, the sensitivity of the devices should be at the sub mPa level.
Also, a good pointing resolution may serve the purpose of UHE point source search. Given the giant array layout introduced above, an acoustic signal will be recorded by a small number of acoustic devices. It is therefore desirable to design acoustic sensor devices with pointing capabilities of their own.


In the next section, we present R\&D activities which are taking place at our lab in regard of sensor design and construction and discuss its performances.

\section{R\&D activities}
The design and construction of a multi-channel sensor was conducted at our lab, which use piezo transducers (PZT) as sensitive elements.
A noise level level $S/N<5$ mPa ($S/N \equiv S_\mathrm{RMS}/N_\mathrm{RMS}$) and a good angular resolution were demonstrated, suggesting the possibility for excellent vertex localization combining the responses from all sensor hits. 
The design, which must still be improved to meet our design goals, could eventually allow  for diffuse acoustic noise reduction through spectral shape analysis and accurate energy estimate of physical events.

The setup for conducting the R\&D activities consists of a bath, topped with a support structure for one absolutely calibrated hydrophone (Sensortech SQ03), one homemade sensor and one transmitter. A datataking LabView program interfaced to a National Instrument board is used for analog response digitization (12 bits $\times$ 12 channels, total 1.25 MHz), transmitter pulse generation and $4\pi$ relative orientation between the transmitter and the sensors for automatized sensor profiling.

\begin{figure}[!h]
  \centering
  \includegraphics[width=3in]{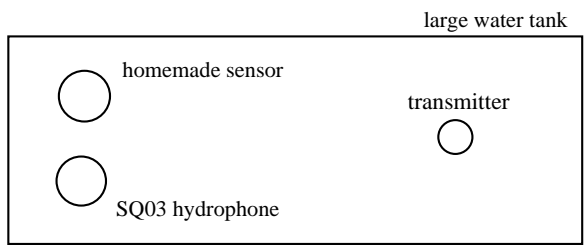}
  \caption{Experimental setup.}
  \label{simp_fig}
 \end{figure}

The experimental setup shown in Fig. \ref{simp_fig} consists of the homemade hydrophone, a transmitter and the reference hydrophone.
In the following, two different electric signal shapes have been considered: a damped sin pulse and a gaussian pulse, resulting in a tripolar pressure pulse (the neutrino-induced thermoacoustic pulse is bipolar). 

\subsection{Acoustic sensor design}

The sensor consists of an aluminium pressure vessel housing 4 channels to provide triangulation capabilities. The noise level at the amplifier input to 130 nV, reached at the expense of some bandwidth reduction, peaking at 22 kHz with $\sim$90 dB amplification and sharply decreasing below 10 kHz and above 40 kHz. Whether that is optimal has to be studied further. It is manufacturable at relatively low cost.
While aluminium is an adequate medium for use in a liquid water bath, it will be replaced by steel for application in ice (more adequate given both impedance and resistance to pressure).


\subsection{Sensitivity calibration}

Signals with peak frequencies in the range 10 - 90 kHz were recorded with a sampling rate of 330 kHz.
A strong frequency correlation between the transmitted pulse and the sensor response was observed.
Due to the finite size of the bath tub and given the sound speed velocity in water, only the first 150 $\mu$s following the pulse arrival time were analysed in what follows to avoid reflexion artefacts.

\begin{figure}[!t]
  \centering
  \includegraphics[width=3in]{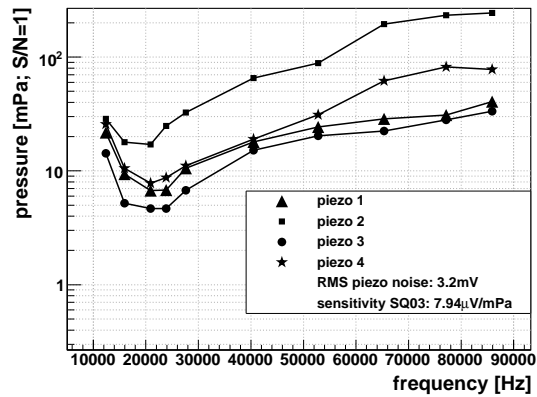}
  \caption{Pressure sensitivity as a function of frequency of damped sin transmitted pulses normalized to $S/N=1$ ratio.}
  \label{fig_sens}
 \end{figure}

With the collected data from the 4-channel sensor and from the commercial hydrophone, the absolute pressure sensitivity was calculated in the time domain using RMS values for signal and noise.
Fig. \ref{fig_sens} shows the absolute pressure sensitivity (defined as $S/N=1$) w.r.t. the dominant frequency of the sent signals. 

The measurements demonstrate the importance of the state of surface coupling the PZT to the housing: The polished surfaces for piezos 1 and 3 show a response $\sim$2 times stronger than piezos 2 and 4 coupled to the housing through porous surfaces.



\subsection{Triangulation}
Time resolution is essential for triangulation and therefore a digitization frequency of 100/200 kHz is required in order to reach cm resolution in aluminium/steel. This suggests that a sensor design should include digital electronics with at least 200 kHz sampling rate per channel\footnote{100 kHz (and therefore 200 kHz sampling rate) is by coincidence the value above which the ice attenuation length drops quickly and roughly the extension of the neutrino-induced thermoacoustic pulse spectrum.}, in order to reach 0.5-5 ms pulse start time resolution (depending on amplitude) which roughly corresponds to 2$^\circ$-20$^\circ$ ($4\pi/10^2$ - $4\pi/10^4$) angular resolution with the current multi-channel sensor. 
The design of a new digital (0.2 MHz/channel) 4-channel amplifier board has been started, with long range communication protocols. It does not yet include trigger logic. Digitization is necessary for a viable acoustic detector design in order to avoid losses in km long cables (of order of 3 dB/100 m in high quality cables) and thus keep both good sensitivity and time resolution.
Once in operation, this will allow to define the requirements for future efficient trigger concept at the sensor level.

 \begin{figure}[!h]
  \centering
  \includegraphics[width=3in]{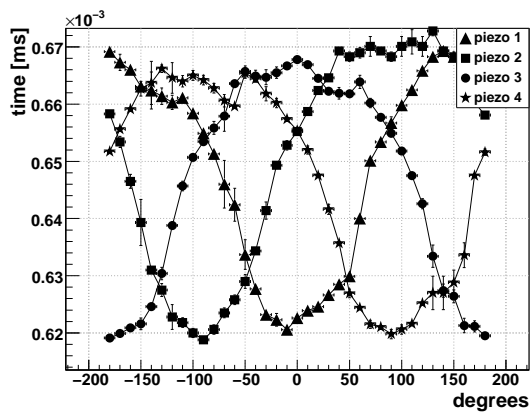}
  \caption{Time of first signal maximum as a function of the polar orientation of the sensor.}
  \label{triang_fig}
 \end{figure}

Figure \ref{triang_fig} demonstrates triangulation capabilities.
Coupling between channels was found to provide in any sampled sensor positions 2 channels with signal within a factor 2 of the channel with the highest response.
The resolution will nevertheless depend on the individual signal-to-noise ratios, but it shows potential for vertex reconstruction with a single sensor module.

\subsection{Outlook}
First positive results in the time domain were obtained. Absolute sensitivities are currently analyzed  in the frequency domain. 
A second 4-channel sensor of similar design in a steel housing will be soon equipped with digital electronics.
Further sensor tests are foreseen to happen next. Low temperature behavior test will be conducted in the laboratory and at large distances and depths in the lake Geneva ($\sim$ 400 m depth) to avoid reflexions, assess the acoustic noise characteristics and probe the sensor design. 

\newpage

\section{Conclusions}
Acoustic neutrino detection techniques should be further developed, pushing the sensitivity at sub mPa levels, together with the characterization of noise sources which may impede the applicability of the technique.
Noise measurement in an open media (water / ice) are required to characterize the noise rate and its spectral shape in order to investigate improved trigger schemes relying on signal processing within the sensors. These developments have been started with digitization board in the sensor, a necessary step for a viable acoustic array design, where signal attenuation along ~km transmission cable is excluded.

While it seems clear that sub mPa sensors should be designed, the uncertain detection conditions at the South Pole make predictions concerning the detection efficiency difficult. The potential can be dangerously spoiled in the case the local source of acoustic noise mimick a neutrino event. Further dedicated studies are on-going to ensure that it could be possible to distinguish the event origin with high efficiency. The deployement of an acoustico-radio-optical hybrid detector would constitute a welcome option, allowing to reduce further possible background noises. Also, complementing the hybrid radio-acoustic strings with a few optical DOMs such as in IceCube would allow to unambiguously tag neutrino events (at these energies, given a $>$100m absorption length in the ice at these depths \cite{ice}, photons may likely accompany a radio-acoustic signal in the case of a neutrino event).

\end{document}